%% file: root.tex
\documentclass[letterpaper, 10pt, conference]{ieeeconf}      % Use this line for a4 paper

\IEEEoverridecommandlockouts                              % This command is only needed if 
\usepackage{amsmath} % assumes amsmath package installed
\usepackage{amssymb}  % assumes amsmath package installed

\usepackage{longtable}
\usepackage[table,xcdraw]{xcolor}
\usepackage{graphicx}

\usepackage{adjustbox}
\usepackage{multirow}
\usepackage{dblfloatfix}
\usepackage{fancyhdr}

\fancypagestyle{firstpage}{%
    \fancyhf{}
    \fancyfoot[l]{\scriptsize This work was accepted by the International Conference on Systems, Man, and Cybernetics, Vienna, Austria, 2025. \textcopyright 2025 IEEE. Personal use of this material is permitted. Permission from IEEE must be obtained for all other uses, in any current or future media, including reprinting/republishing this material for advertising or promotional purposes, creating new collective works, for resale or redistribution to servers or lists, or reuse of any copyrighted component of this work in other works.}
    
}

\title{\LARGE \bf
Conjugated Capabilities: Interrelations of Elementary Human Capabilities and Their Implication on Human-Machine Task Allocation and Capability Testing Procedures
}

\author{Nils Mandischer$^{1}$, Larissa F\"{u}ller$^{1}$, Torsten Alles$^{2}$, Frank Flemisch$^{3}$, and Lars Mikelsons$^{1}$% <-this % stops a space
\thanks{*This work was funded by the Bavarian Hightech-Agenda within the AI Production Network Augsburg.}% <-this % stops a space
\thanks{$^{1}$Chair of Mechatronics, University of Augsburg, 86159 Augsburg, Germany. Corresponding: {\tt\small nils.mandischer@uni-a.de}}%
\thanks{$^{2}$Institute for Quality Assurance in Prevention and Rehabilitation (iqpr), German Sport University Cologne, 50933 Cologne, Germany. Corresponding: {\tt\small alles@iqpr.de}}%
\thanks{$^{3}$Department of Balanced HSI at Fraunhofer FKIE, 53343 Wachtberg, Germany and the Institute of Industrial Engineering \& Ergonomics at RWTH Aachen University, 52062 Aachen, Germany. Corresponding: {\tt\small frank.flemisch@fkie.fraunhofer.de}}%
}

\begin{document}
\bstctlcite{IEEEexample:BSTcontrol}

\maketitle
\thispagestyle{empty}
\pagestyle{empty}

%%%%%%%%%%%%%%%%%%%%%%%%%%%%%%%%%%%%%%%%%%%%%%%%%%%%%%%%%%%%%%%%%%%%%%%%%%%%%%%%
\begin{abstract}
Human and automation capabilities are the foundation of every human-autonomy interaction and interaction pattern. Therefore, machines need to understand the capacity and performance of human doing, and adapt their own behavior, accordingly. In this work, we address the concept of conjugated capabilities, i.e. capabilities that are dependent or interrelated and between which effort can be distributed. These may be used to overcome human limitations, by shifting effort from a deficient to a conjugated capability with performative resources. For example: A limited arm's reach may be compensated by tilting the torso forward. We analyze the interrelation between elementary capabilities within the IMBA standard to uncover potential conjugation, and show evidence in data of post-rehabilitation patients. From the conjugated capabilities, within the example application of stationary manufacturing, we create a network of interrelations. With this graph, a manifold of potential uses is enabled. We showcase the graph's usage in optimizing IMBA test design to accelerate data recordings, and discuss implications of conjugated capabilities on task allocation between the human and an autonomy.
\end{abstract}

%To use machine learning in context of capability-aware machines requires data that is currently unavailable.

%Then, we use combinatorial analysis and rule-based optimization to craft test designs. By this, minimalist tests that cover all relevant quantification of capabilities may be uncovered, that accelerate data collection on human capabilities and performance.

\thispagestyle{firstpage}
%%%%%%%%%%%%%%%%%%%%%%%%%%%%%%%%%%%%%%%%%%%%%%%%%%%%%%%%%%%%%%%%%%%%%%%%%%%%%%%%
\section{Introduction}
\label{sec:intro}
In light of an aging society in most of the leading economies, a key effort within the upcoming years will be to mitigate the consequences of the demographic change, particularly in the working body of companies~\cite{Vaupel.2006}. Human-machine interaction (HMI) is a key technology for enabling people to work in ways that otherwise would not be possible. In the USA, 24.3\% of people aged 65-75 years have a disability~\cite{Elflein.2024}. As people are already projected to work to much older ages within the next decades~\cite{Andrews.2024,Ayuso.2021}, holding the elderly -- and with them their experience -- in work must be a central effort in assistive technologies.

In prior work, we have introduced a capability framework involving capability deltas, that describe limitations in human capabilities, and indicate how a robot or machine shall act in order to mitigate these limitations~\cite{Mandischer.2024c}. While capability deltas are the limitation within individual elementary capabilities (e.g., reaching, gripping, lifting), using these exact limitation scores for interaction planning, gives only one possible solution that is also pretty conservative. In fact, using capability-wise deltas leads to the automation taking over many processes, tasks, and actions that are still actable by the human, as the framework does not make use of reserves in capabilities that are under-challenged. In~\cite{Mandischer.2024c}, we already sketched the idea of the \emph{Delta Compensation Pattern}, that shifts effort from an over-challenged capability (deficit) to an under-challenged capability (usable reserve).

In this paper, we introduce conjugated capabilities, that are interrelated, and which allow to shift efforts on a bilateral basis. For example, if the capability to reach forward is impaired, a person may not reach an object on a table. The capability to reach forward has a deficit. However, if the person can bend the torso, they may make use of this reserve to still drive the hand forward and reach the object. While this pattern is obvious for the human, enabling machines to make use of it is complex. We analyze the IMBA\footnote{From German: ``Integration of people with disabilities into work''.} standard, that defines capabilities, their quantification, and indicators that lead to a diagnosis~\cite{Mandischer.2024c,Mandischer.2024b}. From the standard, pairwise dependent capabilities are deducted that are set into relation (Section~\ref{sec:conjugated}). These relations are also observed in data of rehabilitation patients (Section~\ref{sec:data}). From the conjugated capabilities, we build a network that allows to automatically assess possible pairs or higher-dimensional dependencies within human capabilities needed for the application of stationary manufacturing in standing or sitting posture. This network enables manifold uses in human-machine systems (HMS). We showcase the network in synthesis of minimal-time test plans for data recording (Section~\ref{sec:capability_newtork}), and indicate its implications on task allocation in HMS (Section~\ref{sec:disability}).

\section{Related Work}
\label{sec:relatedwork}

\subsection{Quantified Scales and Capability Assessment}
\label{ssec:scales}
In medical assessment systems, quantified scales are common, particularly in rehabilitation where one of the central tasks is to assess the state of the patient and apply therapy, accordingly. Many items within these scales may be interpreted as capabilities. In stroke rehabilitation, the Fugl Meyer Assessment (FMA-UE/FMA-LE)~\cite{FuglMeyer.1975} defines six sensational and 33 or 17 motor function capabilities for the rehabilitation of upper or lower extremities, respectively. The FMA uses three quantifications $\mathbf{Q}_{FMA}=\{0,1,2\}$ per item, where 0 is the complete lack of the capability. Examples of capabilities in FMA-UE are shoulder retraction, elbow flexion, or pronation/supination, focusing on the arm and shoulder system. The Action Research Arm Test (ARAT)~\cite{Lyle.1981} defines four sub-scales with a total of 19 items quantified by $\mathbf{Q}_{ARAT}=\{0,1,2,3\}$. The sub-scales grasp, grip, pinch and gross movement may be interpreted as capabilities, while the items give nuanced information on the capabilities.
%, hence, the true quantification is the sum of all items within a sub-scale.
The Wisconsin Gait Scale (WGS)~\cite{Rodriquez.1996} defines 14 items with a variable quantification for gait analysis. Only selected items may be interpreted as capabilities, e.g., pelvis rotation or hip extension. The Berg-Balance-Scale~\cite{Berg.1989} defines 14 items with a quantification $\mathbf{Q}_{BBS}=\{0,1,2,3\}$. The majority of items may be interpreted as capabilities, e.g., [movement from] sitting to standing or reaching forward. All these tests are focused on a specific anatomical system, lacking comprehensive capabilities and partially unified quantification.

IMBA~\cite{IMBA.2019} defines \mbox{$n_{main}=70$} capabilities on the main level (e.g., hand/finger motion) and \mbox{$n_{detail}=106$} on the detailed level (e.g., pinch grip), with a quantification \mbox{$\mathbf{Q}_{IMBA}=\{0,1,2,3-,3+,4,5\}$}. The quantification of the capabilities is propagated such that the lowest score in a set of detailed capabilities gives the score of the according main-level capability. The standard is used to assess patients in rehabilitation, and is designed for all possible limitations, which makes it particularly powerful for exploitation in HMS. Schaffernicht, Moder, and Quendler~\cite{Schafferni.2021} use IMBA to match people with disabilities (PwDs) and horticulture work processes, indicating good fit of manual work, PwDs and the assessment through IMBA. H\"{u}sing et al.~\cite{Husing.2021} exploit and extend IMBA for task allocation in human-robot collaboration (HRC) with PwDs, based on manual assessment of capability and process profiles.

Using elementary capabilities as defined by IMBA has significant benefits over skill-based task planning approaches for HMS, e.g., \cite{Gonnermann.2022,Mower.2021,Graler.2021,Schou.2018}. Skills are an abstract agglomeration of elementary capabilities, e.g., ``can manipulate'' or ``can grasp''. Skill-based planning gives agents (humans or robots) attributes in the form of binary skill values, which limit the usability for people with nuanced limitations. Further, as skills are treated as gates to specific tasks, the lack of nuances makes sub-structuring tasks impossible. Thus, the collaboration paradigm is not really applied at all. Instead, tasks are mutually allocated between agents, even though sub-tasks may be performable by the human. This lack of nuancing may be overcome by elementary capabilities, that eventually allow tasks to be performed in a real collaborative manner. Weidemann et al.~\cite{Weidemann.2023b} implement parts of IMBA in PDDL and validate the feasibility of elementary capabilities for task planning, but do not include collaborative actions.

%Hassan and Curry~\cite{Bertino.2013} propose using dummy tasks to measure human capabilities, and then using them for task matching on crowdsourcing platforms.

\subsection{Capability Framework}
In accordance with~\cite{Mandischer.2024c}, we define a set of $n$ capabilities~$c_{j}$, $1\le j\le n$, where each capability has a finite quantification $c_{j}^{i}\in\mathbf{Q}, \mathbf{Q}\subset\mathbb{N}_{0}$, relative to a specific person (or agent) $i$ with their individual capability profile \mbox{$\mathbf{p}^{i}=(c_{j}^{i})_{j\in \mathbf{P}}$}. Hereby, \mbox{$\mathbf{P}=\{1,...,n\}\subset\mathbb{N}$} is the linear index set of $\mathbf{p}$ denoting all potential capabilities $c_{j}$. In the IMBA standard, \mbox{$\mathbf{Q}=\{0,1,2,3,4,5,6\}$}, whereas $0$ is the lowest and $6$ the highest quantification of the capability, and $3$ and $4$ are trivially interpret as sub-populations of value $3$: $3-$ and $3+$. The requirements imposed by an action $k$ (e.g., grasping an object) are defined in the requirements set \mbox{$\mathbf{b}^{k}=(r_{j}^{k})_{j\in \mathbf{B}^{k}}$}, with \mbox{$r_{j}^{k}\in\mathbf{Q}$} and $\mathbf{B}^{k}\subseteq \mathbf{P}$. This underlines the power of elementary capabilities, as in this framework a skill is broken down into the elements of $\mathbf{B}^{k}$. As capabilities and requirements appear pairwise, we may compare both to assess the feasibility of an individual $i$ within the action $k$. To this end, we introduced so-called capability deltas~\cite{Mandischer.2024c}
\begin{equation}
    \label{eq:capability_delta}
    \mathbf{\Delta}^{ik} = (\delta^{ik})_{j\in\mathbf{B}^{k}} = (r_{j}^{k}-c_{j}^{i})_{j\in\mathbf{B}^{k}}.    
\end{equation}
The capability deltas allow on a macro-level to decide if the task shall be performed by agent $i$ and on the micro-level, which specific capabilities prevent the task fulfillment. Note, that $i$ is interchangeable for human or autonomous agents.

\subsection{Delta Compensation}
Capabilities may be distinguished into capacity $c_{j}^{i}$ and performance $\hat{c}_{j}^{i}$. While capacity is the performance limit without further influences, performance is the actual performance as an interplay of the human and their interaction context~\cite{Mandischer.2024b}. The capability delta of the capacity indicates capabilities, in which the human has usable reserves (\mbox{$\delta^{ik}<0$}) or deficits (\mbox{$\delta^{ik}>0$}). If in a set of capabilities involved in an action, deficits are present while equal reserves are available, the effort may be redistributed to the capabilities with reserves, to create a capability delta such that
\begin{equation}
    \sum_{j\in\mathbf{B}^{k}}\max{(\delta_{j}^{ik},0)}=0.
    \label{eq:detla_equilibrium}
\end{equation}
To fulfill Equation~\ref{eq:detla_equilibrium}, compensation measures have to be taken. As shown in the example of grasping a far-out object in Section~\ref{sec:intro}, one measure is to lower the requirement on a deficient capability and simultaneously raise the requirement on an interrelated capability with reserves. By this, tasks that were originally not performable by the human based on the pure assessment of the rather static requirements may become performable. We described this process in the \emph{Delta Compensation Pattern}~\cite{Mandischer.2024c}. For the interrelation of capabilities, we already introduced the term conjugated capabilities. These are detailed out in this work.

\section{Conjugated Capabilities}
\label{sec:conjugated}
We define conjugated capabilities as a pair of capabilities~\mbox{$\langle c_{j_{1}},c_{j_{2}} \rangle$}. For example: To look to the side, either the head, the torso, or both may be moved. In this example, both elementary capabilities are interrelated and build a pair of conjugated capabilities. Even though the majority of medical and occupational assessment standards assume capabilities to be mutually independent to prevent evaluation of the same symptoms within multiple capabilities, they are partially interrelated in application, e.g., as capabilities belong to the same anatomical function or are influenced by the same illness. These interrelations can be used for mining conjugated capabilities.

\iffalse
\begin{equation}
    \label{eq:conjugated_capability}
    \langle c_{j_{1}},c_{j_{2}} \rangle.
\end{equation}
\fi

\subsection{Selection of Capabilities and Restrictions}
While the capability framework is designed to utilize any arbitrary subset of capabilities, we aim at using the framework in manufacturing processes. Hence, for an initial study, we select capabilities that are observable in scenarios at stationary manufacturing workstations. All capabilities must be observable over the worktable (over-table), independent of the worker standing or sitting (e.g., in case of wheelchair). We, further, only consider physical and physiological capabilities, as mental capabilities influence a multitude of other capabilities and are not directly observable by means of visual sensing. In an automated assessment, some capabilities are required to understand the communication of instructions by the machine, e.g., visual acuity, others are not measurable independently of other capabilities by means of observation, e.g., dexterity and perception of motion. While we account for these capabilities as having an influence on almost all other capabilities, we sort them out for separate consideration in an upstream test. For example: If the human shall fulfill a task without external support, they require sufficient sight to perceive the task and deduct what has to be done. If the sight is insufficient, the behavior will change accordingly, hence, it is likely to become impossible to separate the impact of sight and the conjugated capability, e.g., reaching. The upstream tests also feature pushing/pulling as it is not observable through visual observation, but requires specific sensing measures (force sensing). For all bilateral or unilateral capabilities, we only consider the unilateral variant, e.g., in the main capability Hand/Finger Movement. The resulting capabilities are listed in Table~\ref{tab:selected_caps}.
\begin{table}[tb]
    \caption{Capabilities relevant for working at stationary manufacturing workstations.}
    \centering
    \begin{tabular}{l l}
        \hline
        \rowcolor[gray]{0.9}\multicolumn{2}{c}{Upstream Tested Capabilities}\\
        \hline
        1.01 & Sitting \\
        \hline
        1.02 & Standing \\
        \hline
        4.01.01 & Visual Acuity -- Near \\
        4.01.02 & Visual Acuity -- Far \\
        4.01.03 & Depth Perception \\
        4.01.04 & Field of Vision \\
        \hline
        4.04 & Touching/Feeling \\
        \hline
        4.05.01 & Perception of Movement and Position - Upper Extremities \\
        \hline
        5.03.01 & Pushing \\
        5.03.02 & Pulling \\
        \hline
        5.06.02 & Hand Dexterity -- Unilateral\\
        5.06.04 & Finger Dexterity -- Unilateral\\
        \hline
        \rowcolor[gray]{0.9}\multicolumn{2}{c}{Main Over-Table Capabilities}\\
        \hline
        1.05.01 & Sitting -- Bent Posture ($\le$ 30°)\\
        1.05.02 & Sitting -- Stooped Posture ($>$ 30°)\\
        1.05.03 & Standing -- Bent Posture ($\le$ 30°)\\
        1.05.04 & Standing -- Stooped Posture ($>$ 30°)\\
        \hline
        1.06.01 & Sitting/Standing -- Arms Horizontal in Front\\
        1.06.02 & Sitting/Standing -- Arms Over Head\\
        \hline
        3.01.01 & Head/Neck -- Rotation Movements \\
        3.01.02 & Head/Neck -- Bowing \& Stretching \\
        3.01.03 & Head/Neck -- Sideways Bending \\
        \hline
        3.02.01 & Trunk -- Rotation Movement while Sitting \\
        3.02.03 & Trunk -- Bending/Stretching \\
        \hline
        3.03.02 & Reaching Above Head/Shoulder -- Unilateral\\
        3.03.04 & Reaching Forward -- Unilateral\\
        3.03.06 & Reaching Sideways -- Unilateral\\
        3.03.08 & Reaching Backward -- Unilateral\\
        3.03.10 & Turning the Lower Arms -- Unilateral\\
        \hline
        3.04.02 & Hand -- Gripping (Fist Grip) -- Unilateral\\
        3.04.04 & Hand -- Applying Pressure -- Unilateral\\
        3.04.06 & Hand -- Turning -- Unilateral\\
        3.04.08 & Finger -- Pinch Grip -- Unilateral\\
        3.04.10 & Finger -- Applying Pressure -- Unilateral\\
        \hline
        5.01.01 & Lifting -- Horizontal \\
        5.01.03 & Lifting -- Waist to Eye Level \\
        5.01.04 & Lifting -- Waist to Above Head Level \\
    \end{tabular}
    \label{tab:selected_caps}
\end{table}
The nomenclature of IDs is according to~\cite{Mandischer.2024c}: \emph{complex}.\emph{main}.\emph{detail}, e.g., $c_{4.01.01}$ is the first detail (``Near'') of the main capability ``Vision'', which is the first capability of complex ``Information''. In complexes or main capabilities, the following sub-IDs may be omitted.

\subsection{Interrelations of Elementary Capabilities}
\label{ssec:interdependency}
We analyze the interrelations of the elementary capabilities in Table~\ref{tab:selected_caps} by using (a) the description of diagnostics within the IMBA standard~\cite{IMBA.2019}, (b) by considering general anatomy, and (c) through experimental testing. As conjugation of capabilities is mostly equal in sitting or standing posture, we will only discuss sitting postures in the following, omitting $c_{1.02}$, $c_{1.05.03}$, and $c_{1.05.04}$. The resulting interrelations are characterized by following relations:
\begin{itemize}
    \item[d] depends on
    \item[c] condition for
    \item[a] appears in combination with
    \item[r] may be (partially) replaced by
\end{itemize}
We analyze the interplay within manufacturing scenarios. For this, we brainstorm manufacturing processes, and annotate capabilities that appear in interrelation within manufacturing. These are annotated with \emph{(M)}. The full list of characterized interrelations is listed in Table~\ref{tab:interdependence_caps} in the Appendix. The table is to be read from rows to columns, e.g., ``Sitting'' ($c_{1.01}$) is a condition for ``Sitting -- Bent Posture'' ($c_{1.05.01}$). By differentiating between conditions and dependencies, we can later avoid loops within graph representations of Table~\ref{tab:interdependence_caps}. This is important to maintain the ability to perform computations on the conjugated capabilities.

\section{Data on Interrelations}
\label{sec:data}
For validation of the interrelations, we analyze data of IMBA profiles by the Austrian Pension Insurance Institution. The data covers patients during and after rehabilitation. There are 1324 samples at the beginning and 1040 samples at the end of the rehabilitation, whereas all end samples also have a corresponding start sample. The sample population has a mean age of $57.19$ years with a range between $25$ and $68$.
\begin{figure*}[b!]
    \centering
    \includegraphics[width=.98\linewidth]{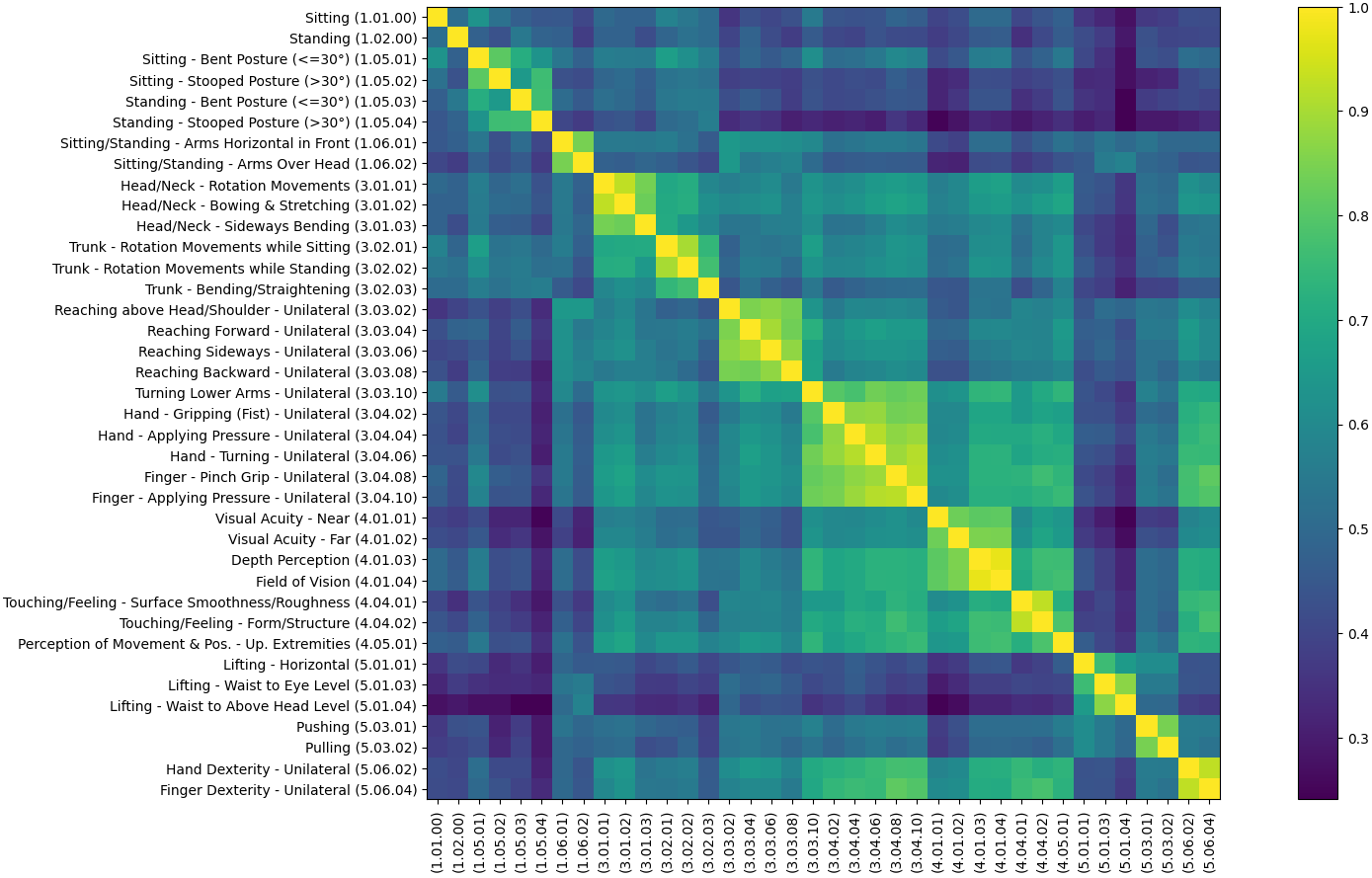}
    \caption{Pearson correlation of relevant capabilities based on $476$ samples of IMBA profiles. Largest power of all capabilities is $p_{max} = 0.0002$.}
    \label{fig:correlation}
\end{figure*}

\subsection{Data Preparation}
While the inter-rater reliability (i.e. raters choosing the same scores for the same condition) of the IMBA standard has never been tested, Achterberg et al.~\cite{Achterberg.2013} observed a good reliance in the related standard MELBA~\cite{MiroGmbH.}. Due to the high overlap of items in both standards, it is likely, that also the inter-rater reliability in IMBA is good. Still, there are significant flaws in the data, that need to be accounted for. Commonly, patients are in rehabilitation for individual illnesses, hence, the majority of capabilities are not relevant for the individual evaluation of the patient. Therefore, therapists often resort to setting all capabilities to the same value or leaving some capabilities empty. To counter this, we filter out all profiles that are incomplete or with a standard deviation $s^{2}<0.2$. Note that profiles centering about the center values $3$ and $4$ are common as IMBA assumes capability profiles to be normal distributed, hence, low standard deviation is expected. We chose $0.2$ by comparing the standard deviation of all samples \mbox{($s^{2}_{post}=[0, 1.707]$, $s^{2}_{post,median}=0.521$)}. For validation, we compute the Pearson correlation between all capabilities of Table~\ref{tab:selected_caps} for sitting posture. To prevent same capabilities to be accounted twice that were not subject to therapy, we use only post-rehabilitation samples. For example, in upper limb rehabilitation, lower limb capabilities are not influenced by the therapy, hence, lower limb capabilities would not change but influence the correlation twice, if all data would be included in the study. We choose the post-rehabilitation samples, as these show less temporary illnesses. After removing profiles with missing entries or low standard deviation, $476$ samples (of $1040$) are used for computing the correlations depicted in Figure~\ref{fig:correlation}.

For hypotheses testing, we use a Monte Carlo-based test with $10,000$ resamples. Monte Carlo simulation can be used as an exact statistical hypothesis test to measure the likeliness of two sub-populations being sampled from the same distribution~\cite{Phipson.2010}. Exact tests give more accurate results in populations with $n<500$ than standard power analysis using Laplacian distributions. The highest power of all capability pairs in Figure~\ref{fig:correlation} is $p_{max} = 0.0002$. Hence, the correlation of all elements is of high significance.

\subsection{Discussion}
Figure~\ref{fig:correlation} shows strong correlations within most main capabilities and moderately strong correlation within complexes. Interestingly, ``Turning Lower Arms'' ($c_{3.03.10}$) correlates stronger with the main-level capability ``Hand/Finger Movements'' ($c_{3.04}$) than its superordinate capability ``Arm Movements'' ($c_{3.03}$). This is an indicator, that the capabilities shall rather be ordered in anatomic functions than systems. Another observation is, that the visual acuity which we initially considered a significant influence to test fulfillment, is less strongly correlated to other capabilities besides visual perception. In fact, tactile perception is more influential. The most influential capabilities -- based on the count of strong correlations -- come from the complexes ``Body Movements'' ($c_{3}$) and ``Information'' ($c_{5}$). Note that even the lowest correlation values are still rather high (\mbox{$r_{mean}=0.521$}). The correlations in Figure~\ref{fig:correlation} do not directly validate the presence of conjugated capabilities, but indicate, that some pairs of capabilities are evaluated dependently in the IMBA standard. Note, that the correlation indicates, that a change in one influences the other capability. In conjugated capabilities, a change in a capability does not necessarily cause a change in the conjugated capability. The dependency in the IMBA standard could be caused by an illness or injury influencing multiple capabilities, e.g., in the same system (limbs, organs, etc.) or by rater bias, or indicate flaws within the standard (double evaluation of same feature). However, as illness or injury influence anatomical systems or functions, and such have been considered for crafting the conjugated capabilities in Table~\ref{tab:interdependence_caps}, we can uncover similar patterns in Figure~\ref{fig:correlation}. In addition, rater bias may come from often seen conjugation within capabilities that are interpreted into patients of similar appearance or condition. Therefore, we consider the correlations as a necessary criterion for the presence of general conjugation, that is not necessarily sufficient. However, within the evaluation of the IMBA standard, the conjugated capabilities derived from the correlations align the automated assessment with the human expert.

\section{Capabilities Network and Test Synthesis}
\label{sec:capability_newtork}
\begin{figure*}[tb]
    \centering
    \includegraphics[width=.82\linewidth]{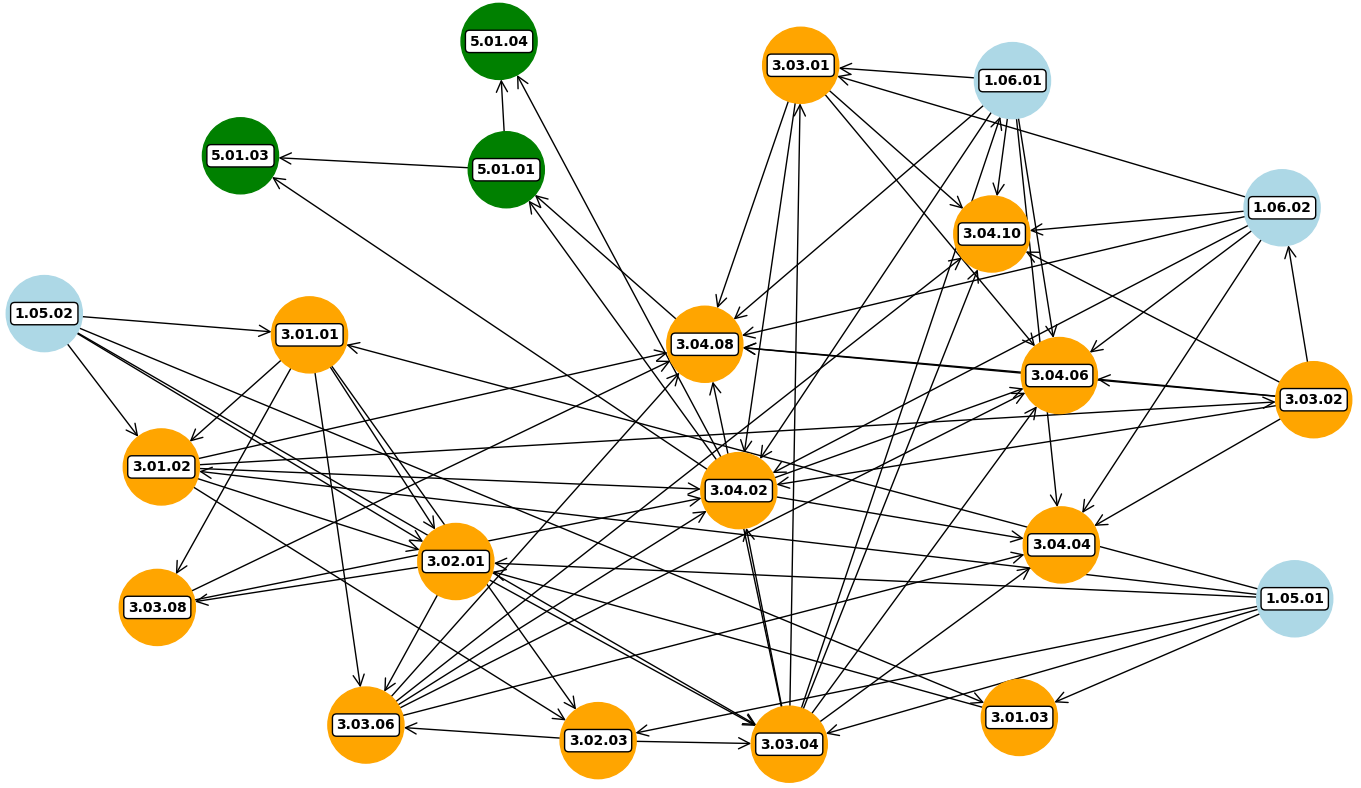}
    \caption{Directed graph of conjugated capabilities according to Table~\ref{tab:interdependence_caps}, adjusted according to the correlation of Figure~\ref{fig:correlation}.}
    \label{fig:graph_caps}
\end{figure*}

In the following, we convert Table~\ref{tab:interdependence_caps} into a directed graph, i.e. a machine-readable form of the conjugated capabilities. The majority of edges connecting two nodes, respectively the conjugated capabilities, have a moderate correlation \mbox{($0.4\le r <0.8$)}. We remove the weak ($r<0.4$) pairs $\langle c_{3.04.02},c_{5.01.04} \rangle$, $\langle c_{3.04.08},c_{5.01.03} \rangle$,$\langle c_{3.04.08},c_{5.01.04} \rangle$, and $\langle c_{5.01.03},c_{3.01.03} \rangle$. Further, we investigate all strongly correlated pairs not in Table~\ref{tab:interdependence_caps} and annotate the feasibility of these candidates in Table~\ref{tab:directed_caps_strong}.
\begin{table}[b]
    \newcommand{\strong}{\cellcolor{gray!50}}       % > .8
    \newcommand{\moderate}{\cellcolor{gray!30}}     % > .4
    \newcommand{\weak}{\cellcolor{gray!10}}         % < .4
    \centering
    \caption{Directed dependencies with strong correlations ($\ge 0.8$), that were not initially uncovered as part of Table~\ref{tab:interdependence_caps}.}
    %\begin{tabular}[t]{c c | c | c}
    \begin{tabular}[t]{p{8mm} p{8mm} | p{5mm} | p{4mm}}
        $c_{j_{1}}$ & $c_{j_{2}}$ & $r$ & type\\
        \hline
        1.05.01     & 1.05.02 & \strong 0.810 & ic \\
        1.06.01     & 1.06.02 & \strong 0.845 & ** \\
        3.01.01     & 3.01.03 & \strong 0.843 & * \\
        3.01.02     & 3.01.03 & \strong 0.831 & * \\
        3.03.02     & 3.03.04 & \strong 0.851 & ** \\
                    & 3.03.06 & \strong 0.869 & ** \\
                    & 3.03.08 & \strong 0.844 & ** \\
        3.03.04     & 3.03.06 & \strong 0.897 & ** \\
                    & 3.03.08 & \strong 0.834 & ** \\
        3.03.06     & 3.03.08 & \strong 0.875 & ** \\
        3.03.10     & 3.04.10 & \strong 0.832 & - \\
        3.04.02     & 3.04.08 & \strong 0.838 & ic \\
                    & 3.04.10 & \strong 0.832 & ic \\
        3.04.04     & 3.04.06 & \strong 0.915 & *** \\
                    & 3.04.08 & \strong 0.870 & ic \\
    \end{tabular}
    \begin{tabular}[t]{p{8mm} p{8mm} | p{5mm} | p{4mm}}
        $c_{j_{1}}$ & $c_{j_{2}}$ & $r$ & type\\
        \hline
        3.04.04     & 3.04.10 & \strong 0.887 & ic \\
        3.04.06     & 3.04.08 & \strong 0.889 & - \\
                    & 3.04.10 & \strong 0.915 & *** \\
        3.04.08     & 5.06.04 & \strong 0.811 & pre \\
        4.01.01     & 4.01.02 & \strong 0.833 & pre \\
                    & 4.01.03 & \strong 0.810 & pre \\
                    & 4.01.04 & \strong 0.811 & pre \\
        4.01.02     & 4.01.03 & \strong 0.850 & pre \\
                    & 4.01.04 & \strong 0.846 & pre \\
        4.01.03     & 4.01.04 & \strong 0.975 & pre \\
        4.04.01     & 4.04.02 & \strong 0.927 & pre \\
        5.01.03     & 5.01.04 & \strong 0.869 & ic \\
        5.03.01     & 5.03.02 & \strong 0.846 & pre \\
        5.06.02     & 5.06.04 & \strong 0.928 & pre \\
        \hline
        3.01.03     & 3.02.01 & \moderate 0.704 & - \\
    \end{tabular}
    \scalebox{.85}{
        \begin{tabular}[b]{l l l l}
            -   & not part of Table~\ref{tab:interdependence_caps} & *   & assumes simultaneous rotations\\
            ic  & impossible & **  & assumes combination of reaching directions\\
            pre & part of pre-test & *** & assumes movement while applying\\
            \ & \ & \ & frontwards pressure
        \end{tabular}
    }
%    \scalebox{.87}{
%        \begin{tabular}[b]{p{4mm} p{70mm}}
%            -   & not part of Table~\ref{tab:interdependence_caps}\\
%            ic  & impossible combination\\
%            pre & part of pre-test\\
%            *   & assumes simultaneous rotations\\
%            **  & assumes combination of reaching directions\\
%            *** & assumes movement while applying frontwards pressure
%        \end{tabular}
%    }
    \label{tab:directed_caps_strong}
\end{table}
Most candidates are infeasible, due to impossible combinations, e.g., simultaneous Pinch Grip ($c_{3.04.02}$) and Fist Grip ($c_{3.04.08}$), or infeasible combinations of movements in the same system, e.g., multiple reaching directions or screw-like motion of the hand. Latter are infeasible, as they happen simultaneously, hence, are hard to distinguish in a test. From this consideration, we add only the two conjugated capabilities $\langle c_{3.03.10},c_{3.04.10} \rangle$ and $\langle c_{3.04.06},c_{3.04.08} \rangle$. As indicated in Section~\ref{ssec:interdependency}, loops are already ruled out by design through the direction of the relations. The resulting network of conjugated capabilities for stationary manufacturing in sitting posture is depicted in Figure~\ref{fig:graph_caps}.
While there are diverse uses for this type of capabilities network in HMS, we will discuss the usage for the synthesis of capabilities test designs in the following. A major barrier in the evaluation of capabilities is the lack of training data. For data recording, unified or standardized testing procedures are required. While there exist tests for at least some of the capabilities in IMBA, these tests are tedious. Our aim is to reduce test durations by synthesizing movement sequences from the graph. Within the movement sequences, conjugated capabilities are shown as direct successors within acted elementary capabilities. To this end, we must make sure that all capabilities in all quantifications are required to be shown by the participant (likely including failures to comply). In post-processing of the tests, combinatorial logic may then be used to isolate individual capabilities and quantifications based on frequency, duration, and failures.

\subsection{Algorithm}
\label{ssec:algorithm}
For minimizing the movement sequences within a test, we formulate an optimization problem. A motion sequence is characterized by a path within the capabilities network, which may begin or end on any downstream node within the graph. Each node shall be visited six times over all paths. We can omit the case of complete lack of a capability, as this is the failure to comply to a quantification \mbox{$c_{j}^{i}=1$}.

Let $x_{w}$ be a binary variable that becomes~$1$ if path $w$ is chosen. Further, let $\eta_{w,c_{j}}$ be a binary variable that becomes~$1$ if path $w$ contains the capability $c_{j}$. The optimization problem is characterized by
\begin{equation}
    \min_{x}{f(x)} = \min{\sum_{w}(x_{w})},
\end{equation}
with the secondary conditions
\begin{subequations}
    \begin{align}
        \sum_{w}{\eta_{w,c_{j}}\cdot x_{w}}& \ge P_{max}, \forall j\in\mathbf{B}^{k}, \\
        \sum_{w}{\eta_{w,c_{j}}\cdot x_{w}}& \le \hat{P}_{max}, \forall j\in\mathbf{B}^{k}.
    \end{align}
\end{subequations}
$P_{max}=6$ denotes the minimum visits per node, and $\hat{P}_{max}$ denotes the maximal tolerable visits per node. Hence, the number of visits per node must be in the range $[P_{max}, \hat{P}_{max}]$. Note that it depends on the structure of the graph, whether a number of exact visits is possible, hence, the usage of a range. First, we compute all paths with minimum length $n_{min}$. This is to prevent a manifold of very short movement sequences. Through experiments, we determined \mbox{$n_{min}=4$} to give aesthetic movement sequences while minimizing the overall paths in the later optimization. Capability ``Head/Neck -- Sideways bending'' ($c_{3.01.03}$) was originally only reachable through the two root nodes ``Sitting -- Bent/Stooped Posture'' ($c_{1.05.01}$ and $c_{1.05.02}$, respectively), with paths of length $n=2$. Hence, the problem becomes infeasible. Therefore, we added another feasible conjugated capability $\langle c_{3.01.03},c_{3.02.01} \rangle$, that has (a) the highest correlation of pairs including $c_{3.01.03}$ (see Table~\ref{tab:directed_caps_strong}) and (b) describes a natural human behavior: Turning trunk and leaning the head, e.g., when looking around an obstacle.

On the set of paths with minimal length, we solve the optimization problem using the COIN-OR Branch and Cut (CBC) algorithm~\cite{JohnForre.2024} and a Simplex solver as part of the Python~3-package \emph{PuLP}~\cite{Mitchell.2009}. Branch and cut is a combinatorial optimization for solving linear programming problems~\cite{Mitchell.2011}. On the nodes of the optimized paths, we annotate requirements $r_{j}^{k}$. We assign quantified requirements from low to high according to first to last encounter of a capability in the paths list. A notable diversion is the capability ``Lifting'' ($c_{5.01}$) which is assigned low values if it is combined with the capability ``Finger -- Pinch Grip'' ($c_{3.04.08}$), and high values if it is combined with ``Hand -- Gripping (Fist)'' ($c_{3.04.02}$), as human's commonly do not use pinch grip for heavy weights but more likely for smaller/lightweight objects.

\subsection{Optimized Test Design}
Optimizing the graph of Figure~\ref{fig:graph_caps} given the criteria in Section~\ref{ssec:algorithm}, gives the 24 movement sequences in Table~\ref{tab:optimized_sequences}.
\begin{table*}[t!]
    \newcommand{\imf}{\cellcolor{gray!75}}
    \newcommand{\ime}{\cellcolor{gray!60}}
    \newcommand{\imd}{\cellcolor{gray!45}}
    \newcommand{\imc}{\cellcolor{gray!30}}
    \newcommand{\imb}{\cellcolor{gray!15}}
    \newcommand{\ima}{}
    \centering
    \caption{Optimized movement sequences. Cells are colored according to requirement, from light (1) to dark (6).}
    \begin{tabular}{r | l | *{10}{m{8mm}}}
        \textbf{ID} & \textbf{Trivial Description} &\multicolumn{10}{|l}{\textbf{Sequence}} \\
        \hline
        0 & pick \& place, from side & 1.05.02\ima & 3.01.03\ima & 3.02.01\ima & 3.03.04\ima & 1.06.01\ima & 3.03.10\ima & 3.04.06\ima & 3.04.08\ima & 5.01.01\ima & 5.01.04\ima \\
        1 & pick \& place, from side & 3.01.01\ima & 3.02.03\ima & 3.03.04\imb & 1.06.01\imb & 3.03.10\imb & 3.04.06\imb & 3.04.08\imb & 5.01.01\imb & 5.01.04\imb \\
        2 & pick \& place, from side & 3.01.03\imb & 3.02.01\imb & 3.03.04\imc & 1.06.01\imc & 3.04.06\imc & 3.04.08\imc & 5.01.01\imc & 5.01.03\ima \\
        3 & pick \& place, from behind & 1.05.02\imb & 3.01.03\imc & 3.02.01\imc & 3.03.06\ima & 3.04.06\imd & 3.04.08\imd & 5.01.01\imd & 5.01.04\imc \\
        4 & seek \& push & 1.05.01\ima & 3.01.02\ima & 3.02.03\imc & 3.03.04\imd & 1.06.01\imd & 3.04.04\ima \\
        5 & pick \& place, from side & 3.02.03\imb & 3.03.06\imb & 3.04.06\ime & 3.04.08\ime & 5.01.01\ime & 5.01.03\imb \\
        6 & pick \& place, from side & 3.03.06\imc & 3.04.06\imf & 3.04.08\imf & 5.01.01\imf & 5.01.03\imc \\
        7 & reach \& push, overhead & 1.05.01\imb & 3.01.02\imb & 3.03.02\ima & 1.06.02\ima & 3.04.10\ima \\
        8 & reach \& push, frontal & 1.05.01\imc & 3.03.04\ime & 1.06.01\ime & 3.03.10\imc & 3.04.10\imb \\
        9 & pull out, from behind & 1.05.01\imd & 3.01.01\imb & 3.03.08\ima & 3.04.02\ima & 5.01.03\imd     & & & & \textbf{Legend}\\
        10 & reach \& push, overhead & 3.01.01\imc & 3.01.02\imc & 3.03.02\imb & 1.06.02\imb & 3.04.04\imb  & & & & $6$\imf\\
        11 & reach \& push, overhead & 3.01.02\imd & 3.03.02\imc & 1.06.02\imc & 3.03.10\imd & 3.04.10\imc  & & & & $5$\ime\\
        12 & pull out, from behind & 3.01.03\imd & 3.02.01\imd & 3.03.08\imb & 3.04.02\imb & 5.01.03\ime    & & & & $4$\imd\\
        13 & pull out, from behind & 3.01.03\ime & 3.02.01\ime & 3.03.08\imc & 3.04.02\imc & 5.01.04\imd    & & & & $3$\imc\\
        14 & reach \& push, frontal & 3.02.03\imd & 3.03.04\imf & 1.06.01\imf & 3.03.10\ime & 3.04.10\imd   & & & & $2$\imb\\
        15 & reach \& push, overhead & 1.05.02\imc & 3.01.02\ime & 3.03.02\imd & 1.06.02\imd & 3.04.04\imc  & & & & $1$\ima\\
        16 & reach \& push, sideways & 1.05.02\imd & 3.01.03\imf & 3.02.01\imf & 3.03.06\imd & 3.04.04\imd \\
        17 & pull out, from behind & 1.05.02\ime & 3.01.01\imd & 3.03.08\imd & 3.04.02\imd & 5.01.03\imf \\
        18 & pull out, from behind & 1.05.02\imf & 3.01.01\ime & 3.03.08\ime & 3.04.02\ime & 5.01.04\ime \\
        19 & reach \& push, sideways & 1.05.01\ime & 3.02.03\ime & 3.03.06\ime & 3.04.04\ime \\
        20 & reach \& push, sideways & 1.05.01\imf & 3.02.03\imf & 3.03.06\imf & 3.04.10\ime \\
        21 & pull out, from behind & 3.01.01\imf & 3.03.08\imf & 3.04.02\imf & 5.01.04\imf \\
        22 & reach \& push, overhead & 3.01.02\imf & 3.03.02\ime & 1.06.02\ime & 3.04.04\imf \\
        23 & reach \& push, overhead & 3.03.02\imf & 1.06.02\imf & 3.03.10\imf & 3.04.10\imf \\
    \end{tabular}
    \label{tab:optimized_sequences}
\end{table*}
Notably, some of the shorter paths are sub-paths of longer paths. We allowed sub-paths, to enable similar test designs for the individual movement sequences. The results give movement sequences that are to be treated as sequence primitives, that may be combined into larger testing procedures, e.g., lifting an object onto a target position and then either (a) extracting the item again or (b) reaching to the item again and lifting it to another -- potentially more challenging -- position. We, further, analyzed the movement sequences in hindsight of human-like behavior and gave them trivial names that summarize the action. Notably, the sequences always start with an alignment of the torso or head with the task object, i.e. establishing sight on the target, followed by reaching towards an object and then performing an action with it (transporting, pushing, or pulling). There is a notable similarity to the elementary cycle of MTM (reach, grip, transport, plug, let loose) and the MTM-Human Work Design (MTM-HWD) process language~\cite{Finsterbus.2016}. From the generated sequences only 17, 18 and 21 are disputable, as the sequences end with an upwards lifting action while arms are in backwards position. This defect can be mitigated by inserting ``Lifting -- Horizontal'' ($c_{5.01.01}$) before the upwards lift or by forbidding the generation of ``Lifting'' ($c_{5.01}$) after ``Reaching Backward'' ($c_{3.03.08}$). Otherwise, we observe human-like motion sequences with no significant flaws within the test design. Hence, this abstract testing procedure is termed a suitable basis to design real tests for data recordings. As a next step, the movement sequences and quantifications must be translated into real-world quantities of human movement and behavior (e.g., reaching distances), together with rehabilitation experts. Further, exact testing procedures need to be crafted from the individual movement sequences, also considering the solubility in combinatorial logic. The translation to an applicable test design may require defining additional partial movement sequences. However, we expect the automated test synthesis to accelerate the design process while indicating new ways of tests, that are not part of conventional rehabilitation medicine. 

\section{Implications on Task Allocation and the Delta Compensation Pattern}
\label{sec:disability}
In real scenarios, the human may temporarily surpass their assessed capability quantifications. Hence, the capability deltas are fuzzy to some degree, and while \mbox{$\sum_{j\in\mathbf{B}^{k}}\max{(\delta_{j}^{ik},0)}>0$} is an indicator that a person is unsuited for an action, this may not be true in application. Therefore, we recommend to add fuzzy parameters as comparators. Note that fuzziness appears on multiple scales: Within individual capability deltas $\delta^{ik}_{j}$, fuzzy parameters $\xi_{j}$ define the maximal deviation possible from the individual quantified $c_{j}^{i}$, with $\xi_{j}\in\mathbb{N}_{0}$ and $|\xi_{j}|\le\max{(\mathbf{Q})}$. Within the set of capability deltas $\Delta^{ik}$, fuzzy parameter $\vartheta$ defines the maximal accumulated over-performance relative to the capacities, with $\vartheta\in\mathbb{N}_{0}$ and $|\vartheta|\le n_{\mathbf{B}^{k}}\cdot\max{(\mathbf{Q})}$, where $n_{\mathbf{B}^{k}}$ is the number of items in $\mathbf{B}^{k}$. Hence, a person is capable to perform a task, if
\begin{subequations}
    \label{eq:fuzziness}
    \begin{align}
        \label{seq:fuzziness_delta}
        \delta^{ik}_{j} &\le\xi_{j}, \forall(j\in\mathbf\mathbf{B}^{k}),\\
        \label{seq:fuzziness_deltaset}
        \sum_{j\in\mathbf{B}^{k}}\max{(\delta^{ik}_{j},0)}&\le\vartheta.
    \end{align}
\end{subequations}
The exact quantification of the fuzzy parameters needs to be calibrated with rehabilitation experts and relative to the context (e.g., work or rehabilitation). Within the \emph{Delta Compensation Pattern}~\cite{Mandischer.2024c}, we briefly discussed another challenge in accordance with the fuzziness: conjugated capabilities. Before we can judge the capability deltas, first, efforts have to be made to shift requirements with $\delta^{ik}_{I}>0$ (control deficit) to capabilities with $\delta^{ik}_{II}<0$ (usable reserve), given the capabilities are conjugated $\langle c_{I}^{i},c_{II}^{i}\rangle$. In~\cite{Mandischer.2024c}, we only sketched this pattern, but now with the fetched out conjugated capabilities, this pattern may be used to its full extend. If a person shall be judged relative to a task, we recommend following procedure:
\begin{enumerate}
    \item Compute capability deltas $\mathbf{\Delta}^{ik}$.
    \item Test capability deltas against Equation~\ref{eq:fuzziness}. If insufficient continue.
    \item If deficient capabilities are part of a conjugated capability pair with a capability with usable reserves, raise the requirement on the conjugated capability and lower the unmet requirement.
    \item Recompute $\mathbf{\Delta}^{ik}$ and repeat steps 2 and 3, until either Equation~\ref{eq:fuzziness} holds or there is no compensation possible according to step 3.
\end{enumerate}
If these steps are successful, a task may be allocated to the person without support by the automation. If no suitable result may be reached, the person is unable to perform. 

\section{Conclusions} 
In this work, we expanded on the concept of conjugated capabilities and the \emph{Delta Compensation Pattern}. First, we introduced the detailed concept of conjugated capabilities and derived candidate capabilities for the application of stationary manufacturing workstations. We examined the candidates for interrelations based on the IMBA standard and anatomy, and characterized them, accordingly. The derived conjugated capabilities were then inspected with data on IMBA profiles. By computing correlations of post-rehabilitation capabilities, we supported the majority of the initial findings, but also uncovered more interrelations. From all conjugated capabilities, we generated a network of capabilities. Finally, we showcased two uses of the capabilities network. First, we showed how the graph representation may be used for synthesis of test plans for rehabilitation assessment and data recordings, by minimizing the test cases required to observe all capabilities in all quantifications. Second, we discussed how conjugated capabilities may be implemented in task allocation, and derived fuzzy task allocation in context of capability deltas. As part of the work context, assistive automation, plays a pivotal role in overcoming occupational performance issues. The proposed conjugated capabilities and patterns are one puzzle piece to shape the use of human-automation systems in context of people with disabilities.

%\addtolength{\textheight}{-8cm}
%\addtolength{\textheight}{-12cm}% This command serves to balance the column lengths

%%%%%%%%%%%%%%%%%%%%%%%%%%%%%%%%%%%%%%%%%%%%%%%%%%%%%%%%%%%%%%%%%%%%%%%%%%%%%%%%

%%%%%%%%%%%%%%%%%%%%%%%%%%%%%%%%%%%%%%%%%%%%%%%%%%%%%%%%%%%%%%%%%%%%%%%%%%%%%%%%

%%%%%%%%%%%%%%%%%%%%%%%%%%%%%%%%%%%%%%%%%%%%%%%%%%%%%%%%%%%%%%%%%%%%%%%%%%%%%%%%
\section*{APPENDIX}
Table~\ref{tab:interdependence_caps}: Interrelations between capabilities of Table~\ref{tab:selected_caps} for sitting postures (see Section~\ref{ssec:interdependency}).
\begin{table*}[t!]
    \centering
    \caption{Interrelations of capabilities relevant for stationary manufacturing workplaces in sitting posture.}
    \label{tab:interdependence_caps}
    \scalebox{0.67}{
        \centering
        \input{table_capabilities}

    }
    \raggedright
    \begin{tabular}{p{20mm} p{20mm} p{20mm} p{35mm} p{35mm}}
        \\
        Legend: & \textbf{d} depends on & \textbf{c} condition for & \textbf{a} apperas in combination with & \textbf{r} may be replaced by
    \end{tabular}
\end{table*}

\section*{ACKNOWLEDGMENT}
We like to thank the Pension Insurance Institution, Department for Applied Research, Innovation and Medical Service Development (Dr. Edlmayer, Mag. Felder) for providing the data used in this research.

%%%%%%%%%%%%%%%%%%%%%%%%%%%%%%%%%%%%%%%%%%%%%%%%%%%%%%%%%%%%%%%%%%%%%%%%%%%%%%%%

\bibliographystyle{IEEEtran}
\bibliography{references}

\end{document}

%% file: table_capabilities.tex
%\begin{tabular}{l|*{31}{c}}
%\begin{tabular}{l|*{31}{>{\centering\arraybackslash}m{4mm}}}
\begin{tabular}{m{8mm}|*{31}{>{\centering\arraybackslash}m{4mm}}}
    ~ & \rotatebox{90}{1.01} & \rotatebox{90}{1.05.01} & \rotatebox{90}{1.05.02} & \rotatebox{90}{1.06.01} & \rotatebox{90}{1.06.02} & \rotatebox{90}{3.01.01} & \rotatebox{90}{3.01.02} & \rotatebox{90}{3.01.03} & \rotatebox{90}{3.02.01} & \rotatebox{90}{3.02.03} & \rotatebox{90}{3.03.02} & \rotatebox{90}{3.03.04} & \rotatebox{90}{3.03.06} & \rotatebox{90}{3.03.08} & \rotatebox{90}{3.03.10} %\\
    & \rotatebox{90}{3.04.02} & \rotatebox{90}{3.04.04} & \rotatebox{90}{3.04.06} & \rotatebox{90}{3.04.08} & \rotatebox{90}{3.04.10} & \rotatebox{90}{4.01} & \rotatebox{90}{4.04} & \rotatebox{90}{4.05.01} & \rotatebox{90}{5.01.01} & \rotatebox{90}{5.01.03} & \rotatebox{90}{5.01.04} & \rotatebox{90}{5.03.01} & \rotatebox{90}{5.03.02} & \rotatebox{90}{5.06.02} & \rotatebox{90}{5.06.04} \\
    \hline
    1.01 & \cellcolor{gray!30} & c & c &  &  &  &  &  & c &  &  &  &  &  &  &  &  &  &  &  &  &  &  &  &  &  &  &  &  & \\
    %checked
    
    1.05.01 & d & \cellcolor{gray!30} & c &  &  & a & a & a & a & c &  & a &  &  & &  &  &  &  &  &  &  &  &  &  &  & a & a &  &  \\
    %checked

    1.05.02 & d & d & \cellcolor{gray!30} &  &  & a & a & a & a & c &  & a &  &  &  &  &  &  &  &  &  &  &  &  &  &  & a & a &  &  \\
    %checked

    1.06.01 &  &  &  & \cellcolor{gray!30} & c &  &  &  &  &  & & d &  &  & a & d & d & d & d & d & d &   &  &  &  d &  & &  & a & a \\
    %checked

    1.06.02 &  &  &  & d & \cellcolor{gray!30} &  & d &  &  &  & d &  &  & & a & d & d & d & d & d & d &  &  &  &  & d &  &  & a & \\
    %checked
    
    3.01.01 &  & a & a & & & \cellcolor{gray!30} & a &  & a &  &  &  & a & a &&  & & &  &  &  &  &  & & &  &  &  &  &  \\
    %checked
    
    3.01.02 &  & a & a &  & c & a & \cellcolor{gray!30} &  & a &  & a &  &  &  & &  &  &  &  &  &  &  &  &  & a & a &  &  &  &  \\
    %checked
    
    3.01.03 &  & a & a &  &  &  &  & \cellcolor{gray!30} &  &  &  &  &  &  & &  &  &  &  &  &  &  &  &  &  &  &  &  &  &  \\
    %checked
    
    3.02.01 & d & a & a &  &  & a & a &  & \cellcolor{gray!30} &  &  & a & r & a & &  &  &  &  &  &  &  &  &  &  &  &  &  &  &  \\
    %checked

    3.02.03 &  & d & d &  &  &  &  &  &  & \cellcolor{gray!30} &  &  &  &  & &  &  &  &  &  &  &  &  &  &  &  &  &  &  & \\
    %checked

    3.03.02 &  &  &  & & c &  & a &  &  &  & \cellcolor{gray!30} &  &  &  & & d(M) & d(M) & d(M) & d(M) & d(M) &  &  &  &  &  & c &  &  &  &  \\
    %checked

    3.03.04 &  & a & a & c &  &  &  &  & a &  &  & \cellcolor{gray!30} &  &  & a & d(M) & d(M) & d(M) & d(M) & d(M) &  &  &  &  &  &  &  &  &  &  \\
    %checked

    3.03.06 &  &  &  &  &  & a &  &  & r &  &  &  & \cellcolor{gray!30} &  & & d(M) & d(M) & d(M) & d(M) & d(M) &  &  &  &  &  &  &  &  &  &  \\
    %checked

    3.03.08 &  &  &  &  &  & a &  &  & a &  &  &  &  & \cellcolor{gray!30} & & d(M) & d(M) & d(M) & d(M) & d(M) &  &  &  &  &  &  &  &  &  &  \\
    %checked

    3.03.10 &  &  &  & a & a &  &  &  &  &  & & a &  &  & \cellcolor{gray!30} & a &  & c & a &  &  &  &  &  &  &  &  &  & c & \\
    %checked

    3.04.02 &  &  &  & c & c &  &  &  &  &  & c(M) & c(M) & c(M) & c(M) & a & \cellcolor{gray!30} & a & a &  &  & d(M) &  &  & c(M) & c(M) & c(M) &  &  & c &  \\
    %checked

    3.04.04 &  &  &  & c & c &  &  &  &  &  &  c(M) & c(M) & c(M) & c(M) & & a & \cellcolor{gray!30} &  &  &  & d(M) &  &  &  &  &  &  &  & a &  \\
    %checked    

    3.04.06 &  &  &  & c & c &  &  &  &  &  & c(M) & c(M) & c(M) & c(M) & d & a &  & \cellcolor{gray!30} &  &  &  &  &  &  &  &  &  &  & c &   \\
    %checked

    3.04.08 &  &  &  & c & c &  &  &  &  &  &  c(M) & c(M) & c(M) & c(M) &  a &  &  &  & \cellcolor{gray!30} &  & d(M) &  &  &  &  &  &  &  &  & c \\
    %checked

    3.04.10 &  &  &  & c & c &  &  &  &  &  &  &  &  &  &  &  &  &  &  & \cellcolor{gray!30} & d(M) &  &  &  &  &  &  &  &  & a \\
    %checked

    4.01 &  &  & & c & c &  &  &  &  &  &  &  &  &  &  & c(M) & c(M) &  & c(M) & c(M) & \cellcolor{gray!30} & r & r &  &  &  &  &  & c & c \\
    %checked

    4.04 &  &  &  &  &  &  &  &  &  &  &  &  &  &  &  &  &  &  &  &  & r & \cellcolor{gray!30} &  &  &  &  &  &  &  &   \\ 
    %checked

    4.05.01 &   &    &   &   &   &  &  &  &  &   &    &     &    &    &  &  & &  &  &  & r &  & \cellcolor{gray!30} &  &  &  &  &  &  & \\
    %checked

    5.01.01 &   &    &   &   &   &  & &  &  &   &    &     &    &    & & d &  &  &  &  &  &  &  & \cellcolor{gray!30} & c & c & d & d &  &  \\
    %checked

    5.01.03 &   &    &   &   &   &  &  &  &  &   &    &     &    &    &  & d &  &  &  &  &  &  &  & d & \cellcolor{gray!30} & c & d & d &  &    \\
    %checked

    5.01.04 &   &    &   &   &   &  &  &  &  &   &    &     &    &    &  & d &  &  &  &  &  &  &  & d & d & \cellcolor{gray!30} & d & d &  &  \\
    %checked

    5.03.01 &   &   &   &   &   &  &  &  &  &   &   &   &   &   & &  &  &  &  &  &  &  &  & c & c & c & \cellcolor{gray!30} &  &  & \\
    %checked

    5.03.02 &   &   &   &   &   &  &  &  &  &   &   &   &   &   & &  &  &  &  &  &  &  &  &  c & c & c & & \cellcolor{gray!30} &  &   \\
    %checked

    5.06.02 & &  &  & a & a &  &  &  &  &  &  &  &  &  & d & d & a & d &  &  & d &  &  &  &  & &  &  & \cellcolor{gray!30} &  \\

    5.06.04 &  &  & & a &  &  &  &  &  &  &  &  &  &  & &  &  &  & d & a & d &  &  &  &  &  &  &  &  & \cellcolor{gray!30} \\
\end{tabular}